\documentclass[aps,showpacs]{revtex4}
\usepackage{graphicx}
\def\no{\noindent}
\def\bc{\begin{center}}
\def\ec{\end{center}}

\def\beq{\begin{equation}}
\def\eeq{\end{equation}}

\def\br{{\bf r}}
\def\bq{{\bf q}}
\def\bk{{\bf k}}
\def\bareta{{\bar\eta}}

\begin{document}

%Title of paper
\title{Dynamical symmetry breaking in a 2D electron gas with a spectral node
\\ % random spinor problem in 2D \\
%{\small (ssb off.tex)}
}

\author{K. Ziegler}
\affiliation{Institut f\"ur Physik, Universit\"at Augsburg\\
D-86135 Augsburg, Germany}
\date{\today}

\begin{abstract}
We study a disordered 2D electron gas with a spectral node in a vicinity of the node.
After identifying the fundamental dynamical symmetries of this system, the spontaneous breaking
of the latter by a Grassmann field is studied within a nonlinear sigma model approach.
This allows us to reduce the average two-particle Green's function to a 
diffusion propagator with a random diffusion coefficient. The latter has non-degenerate saddle points and is 
treated by the conventional self-consistent Born approximation. This leads to a renormalized chemical potential 
and a renormalized diffusion coefficient, where the DC conductivity increases linearly
with the density of quasiparticles. Applied to the special case of Dirac fermions, our approach provides 
a comprehensive description of the minimal conductivity at the Dirac node as well as for the V-shape 
conductivity inside the bands.   
\end{abstract}

\maketitle

%\section*{Extension after referee report}

%-- comparison with WLA, role of the SCBA etc.

%-- Kubo formula for $\mu$ dependent DC conductivity

\section{Introduction}

The prototype of a 2D electron gas with spectral nodes is graphene, where two symmetric electronic bands
created by the underlying honeycomb lattice structure, touch each other at two different points in the Brillouin zone
\cite{novoselov05,zhang05,castro09,abergel10}. 
The surface states of the recently discovered topological insulators is another example for spectral nodes \cite{zhang11}.
It is a remarkable experimental fact that the two-dimensional electron gas in graphene
is always in a metallic state, regardless of its Fermi energy, provided that the sublattice
symmetry of the honeycomb lattice is unbroken. This is particularly
surprising from the theoretical point of view, since the electron gas should be in
a localized state, at least away from the node (Dirac point) \cite{abrahams79}. At the Dirac point,
however, the underlying Hamiltonian has an extra particle-hole symmetry, depending on the type of
disorder though, which may be responsible for metallic (diffusive) behavior.
The experimentally observed metallic state at and away from the node indicates that the diffusive behavior may
not depend on this extra symmetry. We shall discuss in the following that an additional dynamical
symmetry exists which is responsible, regardless of the chemical potential, for a diffusive
behavior. % of a system with pseudo-spinor wave function (i.e. for two electronic bands).

Diffusion can only occur when random scattering is present in the system. This requires some kind of randomness
in the Hamiltonian and the averaging of a physical quantity (e.g., the conductivity) with respect to the random
distribution. A standard method for this procedure is the weak-localization approach (WLA) \cite{altshuler} .
This method has been also applied to graphene \cite{andoetal} with the result that in the presence of only one
Dirac node the system is metallic away from the node. This result has been later questioned, though, 
by Khveshchenko \cite{khveshchenko06}. Unfortunately, the WLA of Ref. \cite{andoetal} provides information 
only about whether or not weak disorder has the tendency to localize. Explicit expressions for the conductivity, 
which could be compared with experiments, are not available. Although the WLA and the
method we shall describe in this paper are based on a weak-scattering expansion, there is a difference in how 
the long-range correlations are taken into account. The WLA uses the summation over maximally crossed diagrams
\cite{langer66}, whereas we will extract massless modes from a spontaneously broken symmetry. 

The paper is organized as follows: After introducing some fundamental quantities for the diffusion in
a quantum system (Sect. \ref{sect:fundamentals}), we discuss the symmetry properties of a two-band model 
in Sect. \ref{sect:structure}. Then we derive an effective nonlinear sigma model for the description of
spontaneous symmetry breaking in our model (Sect. \ref{sect:nlsm}) and its treatment away from the node 
within perturbation theory (Sect. \ref{sect:pt}). This leads us to diffusion due to a massless mode, which 
is discussed and connected to transport properties of graphene in Sect. \ref{sect:discussion}.

\section{quantum dynamics}
\label{sect:fundamentals}

Many characteristic properties of a quantum system, in particular the spectral and large scale properties,
are determined by symmetry properties of the underlying Hamiltonian. However, for the dynamics of a quantum 
system additional symmetries play a role, because the dynamics is not only controlled by the real spectrum but also 
on the complex plane by the advanced and the retarded Green's function $G(\pm i\epsilon)=(H\pm i\epsilon)^{-1}$, 
which are related by Hermitian conjugation:
$
G(-i\epsilon)=G^\dagger(i\epsilon)
$.
This plays a central role in the linear response approach to transport. But since linear response is quite complex
in graphene \cite{rosenstein09}, we will focus here on the transition probability \cite{economou70}
\beq
P_{\br,\br'}(i\epsilon)=K_{\br,\br'}(i\epsilon)/\sum_{\br}K_{\br,\br'}(i\epsilon)
\label{prob0}
\eeq
with
\beq
K_{\br,\br'}(i\epsilon)
%=\langle {\rm Tr}_2\left[ G_{\br,\br'}(i\epsilon)G^\dagger_{\br',\br}(i\epsilon)\right]\rangle_v
=\langle G_{\br,\br'}(i\epsilon)G_{\br',\br}(-i\epsilon)\rangle_v 
\label{a:transition00}
\eeq
and return to the conductivity later.
The average is here with respect to a random variable (e.g. random potential or a random gap) in the
Hamiltonian $H$. Randomness is necessary to provide scattering that breaks translational
invariance.
It is convenient to combine the two Green's functions in the extended Green's function
\beq
{\hat G}(i\epsilon)=\pmatrix{
(H+i\epsilon)^{-1} & 0 \cr
0 & (H-i\epsilon)^{-1} \cr
}
\label{symm_structure}
\eeq
such that with ${\hat H}=diag(H,H)$ we have ${\hat G}(i\epsilon)=({\hat H}+i\epsilon{\hat\sigma}_3)^{-1}$.
Following the standard procedure for disordered systems,
we must replicate this Hamiltonian, either using a fermion-boson pair or $n$ fermion or boson replicas.
Then there is an orthogonal or unitary symmetry which rotates the two-dimensional space that is 
spanned by the two Hamiltonians.
It has been found long time ago that % the corresponding symmetry is
the symmetry breaking due to $\epsilon$ can cause spontaneous symmetry
breaking in the limit $\epsilon\to 0$. The corresponding massless mode leads to a diffusive
behavior \cite{wegner80,efetov97}. 

The definition of (\ref{symm_structure}) was also used as the starting point for Dirac fermions 
with random mass by Bocquet et al. \cite{bocquet00}. Employing a supersymmetric representation,
where ${\hat G}(i\epsilon)$ is applied to a Bose and to a Fermi field,
the gradient expansion of the effective field theory produces an orthosymplectic nonlinear sigma 
model in this case. Unfortunately, the analysis of the latter is quite involved and the transport 
properties cannot be easily extracted.

Prior to the work by Bocquet et al., an alternative approach was suggested by the present author
\cite{ziegler97}
using explicitly the fact that the Dirac Hamiltonian $H=i\sigma_k\partial_k+m\sigma_3$
($\sigma_j$ are Pauli matrices and $\partial_j$ is the spatial (antisymmetric) difference operator 
with $\partial_j\varphi_\br=(\varphi_{\br+a e_j}-\varphi_{\br-a e_j})/a$, $a\sim0$ is the lattice 
constant and $e_j$ is the unit vector in $j$ direction) obeys the relation
\beq
\sigma_1 H^T\sigma_1=-H
\ ,
\label{id0}
\eeq
which constitutes class D according to Ref. \cite{zirnbauer96}. The relation enables us to introduce the structure
\beq
{\hat G}(i\epsilon)=\pmatrix{
(H+i\epsilon)^{-1} & 0 \cr
0 & (H^T+i\epsilon)^{-1} \cr
}
\label{symm_structureb}
\eeq
for the dynamic description.
This choice has a very important advantage over (\ref{symm_structure}), since
the upper and the lower block have the same determinant. Then the upper block 
can act on bosons and the lower block on fermions, providing us with a Bose-Fermi field
theory and a nonlinear sigma model that has only a free massless Fermi (Grassmann) field 
\cite{ziegler97,Ziegler2009}. The latter describes diffusion and gives directly the
experimentally observed minimal conductivity of graphene. Moreover, it reproduces the phase
diagram (one metallic phase and two insulating Hall phases) of 
Refs. \cite{senthil00,chalker01,evers08,beenakker10} 
for a nonzero average mass \cite{Ziegler2009}.

The disadvantage of (\ref{symm_structureb}) over the definition (\ref{symm_structure})
is its restriction to the Dirac point, since a shift by a chemical potential 
$H\to H+\mu\sigma_0$ violates the relation (\ref{id0}). To cure this limitation, we will start 
in the following from (\ref{symm_structureb}) and extend it in such a way that a chemical 
potential can be included \cite{ziegler12}. This will give us a new dynamic structure with a continuous chiral symmetry 
in Bose-Fermi space. The latter can be spontaneously broken and produce a two-component massless 
Fermi (Grassmann) field.

\begin{figure}
\begin{center}
\includegraphics[width=5cm,height=4.5cm]{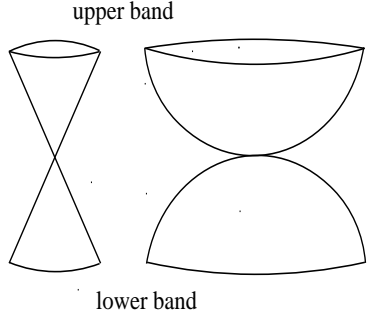}
\caption{
Typical structure of two symmetric bands with a node with linear (left) and quadratic
(right) dispersion.
}
\label{fig:1}
\end{center}
\end{figure}

We consider a Hamiltonian with two bands whose dispersion is symmetric: $\pm E(\bk)$
with the 2D wavevector $\bk$. Moreover, we assume a generalized particle-hole symmetry for the Hamiltonian 
\beq
U H^T U^\dagger =-H , \ \ \ U U^\dagger ={\bf 1}
\ ,
\label{id1}
\eeq
and include a node in the band structure (see Fig. \ref{fig:1}). 
%%%%%%%%%%%%%%%%%%% NEW %%%%%%%%%%%%%%%%%%%%%%%%%%%%%
Besides the Dirac Hamiltonian of Eq. (\ref{id0}), where $H^T\ne H$ and $U=\sigma_1$, this
includes also the symmetric chiral Hamiltonian $H=h_1\sigma_1+h_2\sigma_2$, where $H^T=H$ and $U=\sigma_3$.
An example is the tight-binding Hamiltonian on the honeycomb lattice. The latter consists of two triangular
sublattices A and B, where nearest-neighbor hopping is always between sites on different sublattices.
Thus, the hopping Hamiltonian reads
\beq
H=\pmatrix{
0 &   t_{AB} \cr
t_{BA} & 0 \cr
}
\label{honeycomb}
\eeq 
with the hopping term $t_{AB}$ ($t_{BA}$) from A to B (from B to A). Without a magnetic field the Hamiltonian
is symmetric with $t_{BA}=t_{AB}^T$. This allows us to rewrite the Hamiltonian (\ref{honeycomb}) in the form
of $H=h_1\sigma_1+h_2\sigma_2$ with $h_1=(t_{AB}+t_{BA})/2$ and $h_2=i(t_{AB}-t_{BA})/2$.

%%%%%%%%%%%%%%%%%%% END %%%%%%%%%%%%%%%%%%%%%%%%%%%%%

%, where the Fermi energy is at the node for $\mu=0$. 
The existence of a node is important to create spontaneous symmetry breaking. This
has been observed for gapped Dirac fermions, where the symmetry-breaking solution vanishes
when the gap is too large \cite{Ziegler2009}. 

\section{General Structure and symmetry}
\label{sect:structure}

In analogy to the Green's function in Eq. (\ref{symm_structureb}) we introduce 
${\hat G}(i\epsilon)=({\hat H}+i\epsilon)^{-1}$ with the extended Hamiltonian
\beq
{\hat H}=\pmatrix{
H_+ & 0 & 0 & 0\cr
0 & H_- & 0 & 0\cr
0 & 0 & H_-^T & 0 \cr
0 & 0 & 0 & H_+^T \cr
} , \ \ H_\pm=H\pm\mu\sigma_0
\ .
\label{8by8a}
\eeq
Then, together with property (\ref{id1}), the matrix
\beq
{\hat S}=\pmatrix{
0 & 0 & \varphi_1 U & 0 \cr
0 & 0 & 0 & \varphi_2 U \cr
\varphi_1'U^\dagger & 0 & 0 & 0 \cr
0 & \varphi_2' U^\dagger & 0 & 0\cr
}
\label{symm_tr}
\eeq
with scalar variables $\varphi_j, \varphi_j'$
anticommutes with ${\hat H}$: ${\hat S}{\hat H}=-{\hat H}{\hat S}$. 
This relation implies a non-Abelian chiral symmetry \cite{ziegler97,ziegler12}:
\beq
e^{\hat S}{\hat H}e^{\hat S}={\hat H}
\label{symmetry}
\eeq
which is a symmetry relation for the extended Hamiltonian with respect to ${\hat U}=e^{\hat S}$.

\no
{\it Interpretation of the non-Abelian chiral transformation}:
The anticommuting property ${\hat S}{\hat H}=-{\hat H}{\hat S}$ implies the appearance of two conjugate 
energy bands with energies $\pm E$, respectively, where the eigenstate $|-E\rangle$ %with energy $-E$ 
is created from the eigenstate $|E\rangle$ through the relation ${\hat S}|E\rangle\propto |-E\rangle$ due to
\[
{\hat H}{\hat S}|E\rangle=-{\hat S}{\hat H}|E\rangle=-E{\hat S}|E\rangle
\ .
\]
Iteration of this relation provides the relation ${\hat S}^n|E\rangle\propto |(-1)^nE\rangle$. Thus, $e^{\hat S}|E\rangle$ 
is a linear superposition of states in the upper and lower band $|\pm E\rangle$ with the property
\[
\langle E'|e^{\hat S}{\hat H}e^{\hat S}|E\rangle=\langle E'|{\hat H}|E\rangle
=E\delta_{E',E}
\ ,
\]
i.e., the Hamiltonian matrix is diagonal with respect to the states $e^{\hat S}|E\rangle$, where the diagonal elements 
are the energies. In other words, the variables $\varphi$, $\varphi'$ in ${\hat S}$ create a two-dimensional manifold of
states for which the Hamiltonian matrix is invariant. Moreover, we can define eigenstates to the operator ${\hat S}$ as
$
|\pm S\rangle=|E\rangle\pm |-E\rangle
$, which have the eigenvalues $\pm1$. These eigenvalues characterize the chirality of these states.
%2) The action of ${\hat S}$ does not leave the two-dimensional space spanned by $|\pm E\rangle$. 
%Therefore, the eigenstates of ${\hat S}$ are superpositions of $|\pm E\rangle$, where the two related eigenvalues distinguish
%states with different chirality.
Furthermore, for the eigenstate $|E\rangle$ of ${\hat H}$ the state $e^{\hat S}|E\rangle$ is eigenstate of ${\hat H}^2$ 
with eigenvalue $E^2$ due to
\beq
{\hat H}^2e^{\hat S}|E\rangle=e^{\hat S}e^{-\hat S}{\hat H}e^{-\hat S}e^{\hat S}{\hat H}e^{\hat S}|E\rangle
=e^{\hat S}{\hat H}^2|E\rangle =E^2e^{\hat S}|E\rangle
\ .
\label{adiabatic1}
\eeq
%\no
%4) Evolution: ${\hat H}e^{\hat S}=e^{-{\hat S}}{\hat H}$ and ${\hat H}e^{-{\hat S}}=e^{\hat S}{\hat H}$ implies
%\[
%e^{-i{\hat H}t}e^{{\hat S}}|E\rangle=\left[
%\cos( Et) e^{\hat S}-i\sin(Et )e^{-{\hat S}}\right]|E\rangle
%\]
%which represents a kind  of Rabi oscillation between the two states $e^{{\hat S}}|E\rangle$ and $e^{-{\hat S}}|E\rangle$,
%indicating a simple dynamics between pairs of states. The latter are superpositions of states from the upper band
%and the lower band.  
%\no
%5) two-particle interpretation: The Hamiltonian
%\[
%{\hat H}=\pmatrix{
%H & 0 \cr
%0 & \sigma_1 H^T\sigma_1 \cr
%}
%\]
%represents two independent particles, one with Hamiltonian $H$, the other with Hamiltonian $\sigma_1 H^T\sigma_1$.
%$e^{\hat S}$ mixes the corresponding states without changing the
%Hamiltonian. In other words, the dynamics is invariant under the transformation of the Hamiltonian.
These results indicate that the variables $\varphi$, $\varphi'$ describe an adiabatic change from $|E\rangle$ to $|-E\rangle$.
The corresponding manifold is non-compact, as we can see in the following example of two Dirac fermions with random gap
(cf. Sect. \ref{sect:fundamentals}) and complex $\varphi$ and its complex conjugate $\varphi'$. In this case the matrices read 
\beq
{\hat S}=\pmatrix{
0 & \varphi\sigma_1 \cr
\varphi'\sigma_1 & 0 \cr
}, \ \ \ e^{\hat S}=\pmatrix{
c\sigma_0 & e^{i\phi}s\sigma_1 \cr
e^{-i\phi}s\sigma_1 & c\sigma_0 \cr
}, \ \ \ c=\cosh|\varphi|, \ \ s=\sinh|\varphi| ,\ \ \phi=arg(\varphi)
\eeq
where the variables $c$, $s$ parametrize a non-compact manifold. This is reminiscent of the hyperbolic
saddle-point manifold discovered in the bosonic replica approach to disordered electronic systems, which
is the foundation of the nonlinear sigma model description for Anderson localization  \cite{wegner80}.
%with eigenvalues $\lambda_\pm=c\pm |s|$ and eigenvectors $(1,\pm sign(s) e^{-i\phi})/\sqrt{2}$.

Now we use the matrix structure of (\ref{8by8a}) and apply it to a superspace that
consists of four bosonic (upper) components and four fermionic (lower) components.
In this representation $\varphi_{1,2}$, $\varphi_{1,2}'$ in (\ref{symm_tr}) are Grassmann variables,
and we have to introduce the graded determinant ${\rm detg}$ and the graded trace ${\rm Trg}$ (cf. \cite{Ziegler2009}).
The reason for using a superspace is that for constructing the functional integral of the transition matrix $K$ in 
Eq. (\ref{a:transition00})  it is crucial to have the properties ${\rm detg}({\hat H}+i\epsilon)=1$ and 
${\rm detg}(e^{\hat S})=\exp({\rm Trg}{\hat S})=1$  \cite{ziegler12}. Therefore,
the symmetry is a supersymmetry, connecting bosonic with fermionic degrees of freedom. 
This symmetry is broken by the $\epsilon$ term, though, because ${\hat U}^2$ is not a unit matrix.
Eqs. (\ref{8by8a}), (\ref{symm_tr}) and (\ref{symmetry}) are the main results of this work. What remains to be
discussed is the effect of the symmetry property on the transport % of the 2D electron gas with spectral nodes
for $\epsilon\to0$, which will be studied by the standard nonlinear sigma model approach \cite{coleman,wegner80,efetov97}.

\subsection{ Nonlinear sigma model}
\label{sect:nlsm}

%{\it Nonlinear sigma model}:
The symmetry (\ref{symmetry}) is valid for any random $H$ or $\mu$, provided that $H$ obeys (\ref{id1}).
In order to calculate $K$  of Eq. (\ref{a:transition00}) it is necessary to specify the details of the randomness. 
Once this has been done, it is convenient to employ a transformation from the random variables of the Hamiltonian 
(e.g., random gap or random chemical potential) to the distribution of the diagonal elements of the Green's function
$G_{\br,\br}(\pm i\epsilon)$. The resulting functional integral can be treated within a saddle-point approximation.
%In the following we will 
%consider a random gap because this has a strong effect due to backscattering (unlike a random $\mu$, where
%backscattering is suppressed). For this purpose we consider an uncorrelated Gaussian distributed random gap with 
%mean zero and variance $g$. 
Without repeating here the lengthy and technical but straightforward derivation of the functional integral 
(cf. \cite{Ziegler2009}), we switch directly to the saddle-point approximation of the integral, which allows us to 
focus on the role of the symmetry in Eq. (\ref{symmetry}). 
For this purpose, we start from a special saddle-point solution (which is equivalent to the self-consistent Born approximation 
of the average one-particle Green's function \cite{khveshchenko06}) 
\beq
\langle ({\hat H}+i\epsilon)^{-1}\rangle
\approx (\langle {\hat H}\rangle+i\epsilon+i\eta)^{-1}\equiv{\hat G}_0
\ ,
\label{scba1}
\eeq
where the scattering rate $\eta$ is determined for disorder with strength $g$ by the self-consistent equation
$\eta=2ig {\hat G}_{0,0}$ \cite{ziegler97,ando98}, and perform the functional integration 
only with respect to the symmetry transformation (\ref{symm_tr}), 
independently at each site $\br$. This requires that we replace the two parameters $\varphi_j$ ($j=1,2$) by the space-dependent 
Grassmann fields $\varphi_{j\br}$ such that the integration reduces to the invariant measure, defined through the Jacobian
\beq
J={\rm detg}\left( \langle{\hat H}\rangle+i\epsilon +i\eta {\hat U}^2  %e^{2{\hat S}
\right)^{-1} %= {\rm detg}\left( \langle{\hat H}\rangle+i\epsilon +i\eta {\hat U}^2
\ .
\label{action3}
\eeq
%The parameter $\eta$ is determined by a self-consistent (saddle-point) equation \cite{ziegler97,ando98}. %, as described in App. \ref{app:scba}.
% with the saddle point ${\hat Q}_0=i\eta{\hat \tau}_3$, where the shift of the chemical potential $\mu$ 
% has been included in $ {\hat H}_0$ as $\mu\to{\bar\mu}$.
Thus $K$ of Eq. (\ref{a:transition00}) reads % can be approximated by a functional integral as 
\beq
K_{\br\br'}\approx 4\frac{\eta^2}{g^2}\sum_{j=1,2}\int \varphi_{j\br} \varphi'_{j\br'}J% e^{-S'}
{\cal D}[{\varphi,\varphi'}]
\ ,
\label{corra}
\eeq
where we have summed over $j$, since the diffusion is the same for $\pm\mu$.

Next we expand $-\log J$ in powers of the scattering rate $\eta$. This is also an expansion in powers of ${\hat G}_0$,
as defined in Eq. (\ref{scba1}),
%=({\hat H}_0+i(\epsilon+\eta){\hat \tau}_0)^{-1}$, 
% where the latter can be approximated by a gradient operator. 
which is convergent on large scales \cite{ziegler12}.
%at least on large scales, when the gradient operator is small in comparison to $\eta$ \cite{ziegler12}. 
Up to second order in $\eta$ it reads $-\log J=S''+o(\eta^3)$ with
\[
S''=i\eta{\rm Trg}\left({\hat G}_0{\hat U}^2\right)
-\frac{\eta^2}{2}{\rm Trg}\left[\left({\hat G}_0{\hat U}^2\right)^2\right]
\ .
\]
This is the nonlinear sigma model for the nonlinear field ${\hat U}^2=e^{2\hat S}$. Using the definition of the
latter field in Sect. \ref{sect:structure}, this can also be expressed by the field ${\hat S}$ as
\beq
S''=4i\eta{\rm Trg}\left({\hat G}_0{\hat S}^2\right)
-8\eta^2 {\rm Trg}\left[\left({\hat G}_0{\hat S}\right)^2\right]
-8\eta^2 {\rm Trg}\left[\left({\hat G}_0{\hat S}^2\right)^2\right]
\ ,
\label{expansion2}
\eeq
where the off-diagonal parts of the last two terms give the standard form of the nonlinear sigma model \cite{coleman},
and the first term and the diagonal part of the second term contribute to the symmetry-breaking term that is 
proportional to $\epsilon$. Evaluating the three expansion terms (cf. App. \ref{app:nlsm}) leads to
$S''$ which separates into two components as $S''=S_1''+S_2''$ with
\beq
S_j''=\frac{4\eta}{g}\sum_{\br}\left[
\varphi_{j\br}(\epsilon -D\partial^2)\varphi_{j\br}'
+\alpha_j\Phi_{j\br}\partial^2\Phi_{j\br}\right]
\ ,
\label{action4}
\eeq
with the composite field $\Phi_{j\br}=\varphi_{j\br}\varphi_{j\br}'$ and with the Laplacian $\partial^2=\partial_1^2+\partial_2^2$.
With the Green's function $g_\pm=[\langle H\rangle+i(\epsilon+\eta)\pm{\bar\mu}]^{-1}$ the parameters read for $\epsilon\sim0$
\beq
\alpha=-\frac{\eta^2}{2}Tr_2(g_{+,0}^2-g_{-,0}^2)
=-i\eta^2ImTr_2(g_{+,0}^2)
\ ,
\label{coeff1}
\eeq
$\alpha_j =-(-1)^j\alpha$ and an isotropic diffusion coefficient
\beq
D=-\frac{g\eta}{2}\frac{\partial^2}{\partial q_1^2}\int_\bk Tr_2[{\tilde g}_{+,\bk}(i\eta){\tilde g}_{+,\bk-\bq}(-i\eta)]
\Big|_{\bq=0}
\ ,
\label{coeff2}
\eeq
where ${\tilde g}_{\pm,\bk}$ are the Fourier components of the Green's function $g_\pm$ and $Tr_2$ is the trace with respect to
Pauli matrices.
Thus our model depends only on the parameters $g$ (disorder strength),
$\eta$ (scattering rate), and the renormalized chemical potential ${\bar\mu}$. 
Interestingly, $\alpha$ vanishes for ${\bar\mu}=0$ such that the interaction disappears at the node 
(cf. \cite{ziegler97}).  This enables us to employ an expansion of $J\approx \exp(-S'')$ in powers of $\alpha$
to study the behavior of the integral (\ref{corra}) away from the node. 

According to (\ref{8by8a}) and (\ref{symm_tr}), different values of $j$ refer to different Fermi energies: 
$(H+i\epsilon+\mu)^{-1}$ (for $j=1$) and  $(H+i\epsilon-\mu)^{-1}$ (for $j=2$).
The different signs in front of $\alpha$ in Eq. (\ref{action4}) reflect the fact that $\alpha$ is proportional 
to $\mu$. In the following we ignore the index $j$ because its value affects only the sign of the coupling constant.

\subsection{Perturbation theory around the node}
\label{sect:pt}

At $\alpha=0$ the unperturbed integral simply reads as the adjugate of $\epsilon -D\partial^2$:
\beq
K_{\br,\br'}=\det(\epsilon -D\partial^2)(\epsilon -D\partial^2)^{-1}_{\br,\br'}\equiv Adj_{\br,\br'}(\epsilon -D\partial^2) 
\ .
\eeq
The perturbation expansion on the real space $\Lambda$ then becomes (cf. App. \ref{app:c})
\beq
K_{\br,\br'}=
\sum_I Adj^I_{\br,\br'}(\epsilon -D\partial^2){\prod}'_{\br'',\br'''\in\Lambda\backslash I}\alpha_j\partial^2_{\br'',\br'''}
\ ,
\label{pert_exp}
\eeq
where $Adj^I_{\br,\br'}(\epsilon -D\partial^2)$ is the adjugate on the subspace $I\subseteq\Lambda$, the perturbation is
on its complement $\Lambda\backslash I$, and the summation is over all subspaces $I$.
 
\begin{figure}[h]
\begin{center}
\includegraphics[width=7cm,height=4cm]{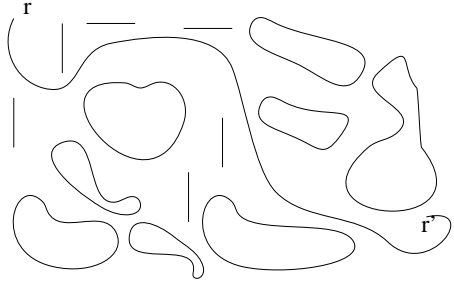}
\caption{
A schematic representation of the expansion in Eq. (\ref{pert_exp}). The random walk of an electron connecting the sites 
$\br$ and $\br'$ and the loops come from the adjugate, whereas the dimers originate from the expansion terms
around the node.
}
\label{fig:dimers}
\end{center}
\end{figure}

%%%%%%%%%%%%%%%%%%%%%%%%%%%%%%%%%%%%%%%%%%%%%%%%%%%%%
The expansion in Eq. (\ref{pert_exp}) can also be understood as a random-walk expansion, where the
perturbation $\alpha_j\Phi_{j\br}\partial^2\Phi_{j\br}$ represent a ``dimer'' on the lattice which cannot be 
visited by the random walk (diffusive path) because of the Grassmann variables. In other words, the perturbation
blocks dimers on the lattice which are not accessible for the random walk, and the actual walk takes place only on the
sites that are not blocked by dimers (cf. App. \ref{app:c} and Fig. \ref{fig:dimers}).
This restricts the random walk of the electron %(cf. Fig. \ref{fig:walk}). 
but since the walk has no phase factor (it is a classical random walk since $\epsilon-D\partial^2$ is a real 
symmetric matrix, as explained in App.  \ref{app:c}), there is no interference to generate Anderson localization. 

Here it should be noticed that a hopping expansion of the two-particle Green's function (\ref{a:transition00})
would also give a random-walk expansion but with random phase terms. This reflects the quantum character of our system.
In particular, the random phase fluctuations can lead to cancellations of expansion terms which may eventually
cause Anderson localization. Unfortunately, such an expansion is difficult to control. The approach of this 
article, in which we have extracted the behavior on large scales in the form of a nonlinear sigma model, allows 
us to connect the system away from the node with the system at the node $\mu=0$ by a classical random walk. Since
no interference appears in classical random walks, our extraction of the massless modes simplifies the calculation
substantially.

\begin{figure}[h]
\begin{center}
\includegraphics[width=8cm,height=7.5cm]{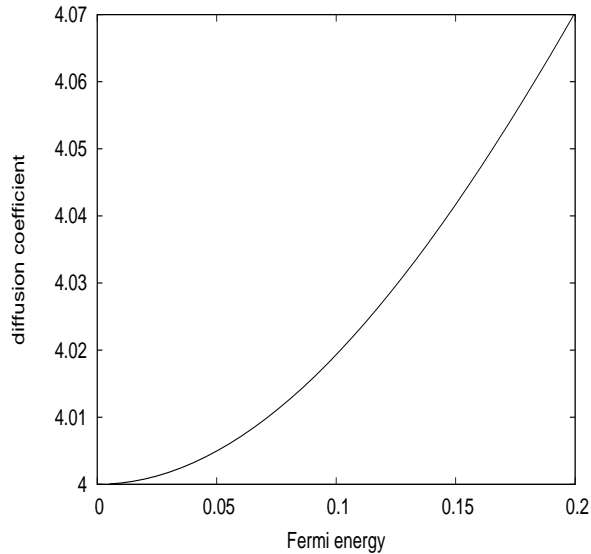}
\caption{
Diffusion coefficient for 2D Dirac fermions (in arbitray units) as a function of the 
renormalized chemical potential ${\bar\mu}$ for $g=1$ (from Eq. (\ref{diff_coeff_final})).
%for the scattering rate $\eta=0.08$ (full curve) and $\eta=0.1$ (dashed curve). The right panel represents the 
%The conductivity in units of $e^2/\pi h$ as a function of the Fermi energy $\mu$ for $g=1$ (from Eq. (\ref{cond1})).
}
\label{fig:2}
\end{center}
\end{figure}

In Ref. \cite{sinner12} we have applied a renormalization-group procedure directly to the action (\ref{action4})
and found that the interaction term scales to zero on large scales. In the following we employ the perturbation
expansion of Eq. (\ref{pert_exp}) as an alternative
approach which will lead us to a similar result. It is based on the idea that the interaction 
can be treated within the self-consistent Born approximation \cite{ando98,andoetal} by replacing the diffusion
coefficient at the node $D$ as $D\to {\bar D}_j= D+(-1)^jD'$. This can be understood as a partial summation of our 
expansion in Eq. (\ref{pert_exp}), where $(-1)^jD'$ is a self-energy.
This approximation should be reliable, since there is no continuous degeneracy of the self-consistent solutions,
in contrast to the supersymmetric functional integral of Eq. (\ref{corra}).
In the special case of
the Dirac Hamiltonian $\langle H\rangle=i\sigma_k \partial_k$ (valid for a single node in graphene or for the
surface of a topological insulator) we get from Eq. (\ref{coeff2}), after performing the $\bk$ integral, the expression
\cite{bernad10}
\beq
D=\frac{g}{8\pi\eta}\left[1+\frac{1+\zeta^2}{\zeta}\arctan\zeta\right] \ ,
\ \ \ \zeta={\bar\mu}/\eta
\ .
\label{diff_coeff_final}
\eeq
For given $g$ and $\mu$ the renormalized parameters $\eta$ and ${\bar\mu}$ are determined as a solution of 
a self-consistent approach (cf. App. \ref{app:b}) with
%${\bar D}_j= D+(-1)^jD'$ with
\[
D'\sim\alpha\lambda^2/4\pi D ,\ \ \ \alpha\sim -i\frac{\bar\mu}{\eta(1+\eta^2/\lambda^2)}
\ ,
\]
where $\lambda$ is the momentum cut-off. 
The Fourier components of $P_\br\equiv P_{\br0}$ in Eq. (\ref{prob0}) then read
\beq
{\tilde P}_\bq={\tilde K}_\bq/{\tilde K}_0=\frac{\epsilon(\epsilon +Dq^2)}{(\epsilon +Dq^2)^2+{D'}^2q^4}
\eeq
and the diffusive expansion of the wavefunction $\sum_\br r_k^2 P_\br$ is
\beq
-{\tilde K}''_{\bq=0}/{\tilde K}_{\bq=0}=2D/\epsilon
\ .
\eeq
It should be noticed that the correction $D'$ drops out and only the diffusion coefficient $D$ enters the final result.
As a result, $D$ is plotted in Fig. \ref{fig:2}. % for two different scattering rates.

\begin{figure}[h]
\begin{center}
\includegraphics[width=8cm,height=7.5cm]{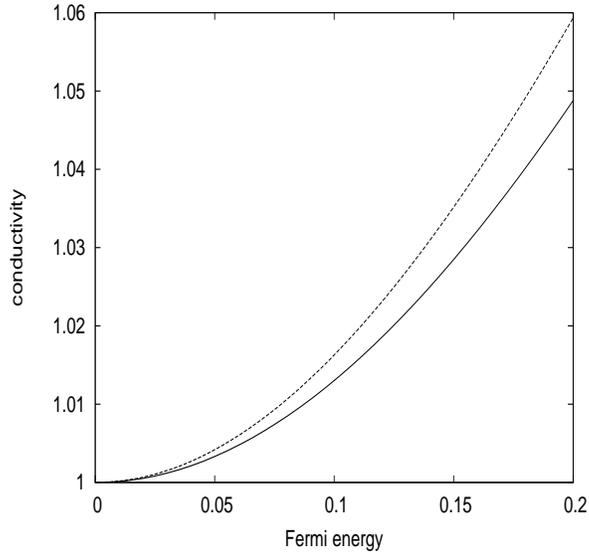}
\caption{
%Diffusion coefficient for 2D Dirac fermions (in arbitray units) as a function of the renormalized chemical potential ${\bar\mu}$
%for the scattering rate $\eta=0.08$ (full curve) and $\eta=0.1$ (dashed curve). The right panel represents the 
The conductivity in units of $e^2/\pi h$ as a function of the Fermi energy $\mu$ (from Eq. (\ref{cond1}))
$g=0.8$ (dashed curve) and for $g=1$ (full curve).}
\label{fig:3}
\end{center}
\end{figure}

Although the dynamic conductivity is quite complex for Dirac particles \cite{rosenstein09}, the 
DC conductivity can be extracted % from $K$ either via the Kubo formula \cite{Ziegler2009} or 
from the Einstein relation as
\beq
\sigma_{kk}\propto \rho D\frac{e^2}{h}
\label{einstein}
\eeq
with the density of states at the Fermi level $\rho$. In fact, the Kubo conductivity \cite{Ziegler2009} gives
\beq
\sigma_{kk}(\mu,0)=4\frac{e^2}{h}\frac{\eta}{g}D %(\mu/\eta,\eta)
=\frac{e^2}{2\pi h}\left[ 1+\frac{1+\zeta^2}{\zeta}\arctan\zeta\right]
\ ,
\label{cond1}
\eeq
where $\zeta$ is a function of the Fermi energy $\mu$ and can be calculated self-consistently 
(cf. App. \ref{app:b}). The result is plotted in Fig. \ref{fig:3}. 
The conductivity increases with $|{\bar\mu}|^2$. This, on the other hand, is proportional to the
quasiparticle density $n$. Consequently, the conductivity increases linearly with $n$. 
This behavior is in agreement with the experimental observation of a V-shaped conductivity
\cite{novoselov05,zhang05,chen08,andrei12}.
In the pure limit without scattering ($\eta \to 0$) we get from Eq. (\ref{diff_coeff_final}) for
the diffusion coefficient $D\to \infty$ and from Eq. (\ref{cond1}) for the conductivity  
\[
\sigma_{kk}\to \cases{
e^2/\pi h & for $\mu=0$ \cr
\infty & for  $\mu\ne 0$ \cr
}
\ .
\]
Thus, at the Dirac node the quantum fluctuations are sufficient to create a finite conductivity,
while away from the node scattering is necessary to obtain a finite conductivity. 

\section{Discussion}
\label{sect:discussion}

%-- comparison with WLA

%-- role of SCBA and factorization of the average 2PGF

The calculation of transport properties near the Dirac node (e.g. in graphene) has been discussed
at length in the literature, using the weak-localization approximation (WLA) \cite{andoetal}. In order 
to understand the connection to our result, in particular in terms of the conductivity in Eq. (\ref{cond1}), 
we briefly survey the Kubo approach within our symmetry-breaking picture and compare it with the WLA.

Starting from the Kubo formula \cite{altshuler,Ziegler2009}
\beq
\sigma_{kk}=-\frac{e^2}{2h}\omega^2\lim_{\epsilon\to0}
Re\left\{\sum_\br r_k^2Tr_2\left[\langle G_{0\br}(\omega/2+i\epsilon) G_{\br0}(-\omega/2-i\epsilon)\rangle\right]
\right\} 
%, \ \ \ G(z)=(H-z)^{-1}
\ ,
\label{cond00}
\eeq
we could employ an approximation for the average two-particle Green's function by
factorizing the averaged product as \cite{altshuler}
\beq
%\sum_\br r_k^2Tr_2\left[
\langle G_{0\br}(y) G_{\br0}(-y)\rangle %\right]
\approx
%\sum_\br r_k^2Tr_2\left[
\langle G_{0\br}(y)\rangle\langle G_{\br0}(-y)\rangle %\right]
\ \ \ (y=\omega/2+i\epsilon)
\ ,
\label{factor0}
\eeq
which neglects the correlations between the two Green's functions at $\pm y$. 
This can be combined with the self-consistent Born approximation in Eq. (\ref{scba1}) to obtain
\beq
%\sum_\br r_k^2Tr_2\left[
\langle G_{0\br}(y)\rangle \langle G_{\br0}(-y)\rangle %\right]
\approx 
%\sum_\br r_k^2Tr_2\left[
G_{0,\br}(y+i\eta)G_{0,-\br}(-y-i\eta) %\right]
\ .
\label{drude0}
\eeq
This gives the leading term of the WLA, namely the Boltzmann (or Drude) conductivity \cite{altshuler,andoetal}.
However, this approximation ignores the long-range correlations of the average two-particle
Green's function that is required for diffusion, since $G_{0,\br}(y+i\eta)$ decays exponentially 
on the scale $1/\eta$. Consequently, the sum on the right-hand side of Eq. (\ref{cond00}) is finite. 
Now the main result of the chiral symmetry-breaking approach (CSBA) is that it extracts the long-range
behavior of the average two-particle Green's function in form of the massless mode, described approximately
by the nonlinear sigma model (\ref{action4}). This implies that the approximations in Eqs. (\ref{factor0}), 
(\ref{drude0}) are replaced by the relation \cite{Ziegler2009}
\beq
\sum_\br r_k^2Tr_2\left[\langle G_{0\br}(y) G_{\br0}(-y)\rangle\right]
%=f(y)\sum_rr_k^2 Tr_2\left[G_{0,r0}(iz(y))G_{0,0r}(iz(-y))\right]
%=f(\eta/y)
=(1+i\eta/y)^2\sum_\br r_k^2 Tr_2\left[G_{0,\br}(y+i\eta)G_{0,-\br}(-y-i\eta)\right]
\ ,
\label{scaling_gf2}
\eeq
where the coefficient on the right-hand side represents correlations of the fluctuating  Green's functions
on large scales. These correlations are negligible only for $\omega\tau\gg 1$ 
($\tau=\hbar/\eta$ is the scattering time). Inserting relation (\ref{scaling_gf2}) into the conductivity, we obtain a 
correction term proportional to 
%$(\omega+2i\eta)^2-\omega^2$
$\eta^2$ in comparison with the factorization approximation.
An alternative way to evaluate these corrections for the conductivity consists of the summation of certain
diagrams from the perturbation theory. An example are maximally crossed diagrams \cite{langer66}, which
is the foundation of the WLA. 
%-- cut-off by frequency $L_\omega=v/\omega$
In Table \ref{comparison} we compare several physical quantities for a single Dirac node 
(i.e., for the long-range potential in the notion of Ref. \cite{andoetal}), 
either calculated in weak-localization approximation (WLA) or in our CSBA. % (or nonlinear-sigma model approach). 
The WLA requires a phenomenological cut-off on the length scale $l_\phi$ that is justified by inelastic 
scattering. Such a cut-off does not appear in the CSBA due to the absence of logarithmic singularities. 

The results of both approaches agree well on a qualitative level for $\mu\gg 0$: $D\propto \tau$, 
$\sigma\approx const.$ and 
$\tau\propto \mu^{-1}$. On the other hand, for small Fermi energies close to the Dirac point, where the 
WLA is not reliable \cite{andoetal}, there is a significant deviation. In this regime the CSBA approach should
also be reliable, as long as disorder is not too strong. It gives $D\approx g\tau/4\pi$, 
$\sigma\approx e^2/\pi h$ and $\tau\approx e^{v^2/g}$ ($v$ is the Fermi velocity).
The disagreement between the results of the two approaches near the Dirac node can be explained by the fact 
that the WLA is based on choosing a special subclass of diagrams of the perturbation theory. For 2D Dirac fermions,
however, we have seen that there is a cancellation of the logarithmically divergent diagrams in each order if
all diagrams are considered \cite{sinner11}. In contrast to the WLA, the CSBA projects onto the chiral symmetry-breaking
modes first and then uses an expansion in powers of $\eta=\hbar/\tau$ in the exponent of the Jacobian.
%in Eq. (\ref{expansion2}). 
This may explain the different results for the two approaches in Table \ref{comparison}.

Both approaches involve certain approximations which cannot be justified directly because they are based
on the truncation of a perturbation series. For instance, the CSBA neglects the massive (short-range correlated) 
fluctuations of the Green's functions, and the WLA takes into account only the maximally crossed diagrams.
Thus, although both approaches are systematic, they lead to different results. They can be understood
as new well-defined models for transport, describing different physical situations. This brings us in the position to
compare their respective results with experimental observations. Transport properties of disordered 
two-dimensional Dirac fermions have been studied intensively in the case of graphene. The conductivity
is characterized by its robust minimal value at the Dirac node $\mu=0$ with $\sigma_{min}\approx 4e^2/h$
(the factor 4 is the result of a four-fold degeneracy due to the spin and two valleys).
Away from the Dirac node the conductivity increases with $\mu^2$ \cite{novoselov05} and with
a prefactor that decreases with the disorder strength \cite{chen08}. On the other hand, the theoretical results in 
Table \ref{comparison} give an infinite negative minimal conductivity for the WLA and $\sigma_{min}=4e^2/\pi h$ 
(taking into account the four-fold degeneracy) for the CSBA.
Besides the factor $1/\pi$, the CSBA agrees quite well with the experiments. It should be noticed though that the WLA
is not reliable near the Dirac node, as mentioned above. Away from the Dirac node the WLA gives a conductivity that
increases logarithmically with the Fermi energy $\mu$, in contrast to the parabolic increase of the CSBA result in 
Fig. \ref{fig:3}. 
Moreover, the WLA correction of the conductivity increases logarithmically with disorder strength $g$, 
whereas the CSBA conductivity decreases with increasing $g$ (cf. Fig. \ref{fig:3}).
These results indicate clearly that the CSBA agrees much better with the experimental observations.  

\begin{table}
\begin{center}
\begin{tabular}{ccc}
 & WLA & CSBA \\
diffusion coefficient $D$ & $v^2\tau$ & $\frac{g\tau}{8\pi}\left[1+\frac{1+\zeta^2}{\zeta}\arctan\zeta\right]$ \\
conductivity $\sigma$ & $\frac{8e^2}{h g\tau}D+\frac{4e^2}{\pi h}\log(l_\phi/2v\tau)$ & $\frac{4e^2}{h g\tau}D$ %$\frac{e^2}{2\pi h}\left[ 1+\frac{1}{\mu\tau}\left(1+\mu^2\tau^2\right)\arctan(\mu\tau)\right]$ 
\\
scattering time $\tau$ & $2\hbar v^2/g\mu$ & $t(\mu) e^{v^2/g}$ \\
\end{tabular}
\caption[smallcaption]{Comparison of the WLA \cite{andoetal} and the CSBA for physical quantities of
a single Dirac node. $g$ is measured in units of the squared Fermi velocity $v^2$ and we have used
$\zeta={\bar\mu}\tau/\hbar$. The function $t(\mu)$ can be taken from Eq. (\ref{parameters}) and must be calculated 
from the self-consistent equation (\ref{spe3}). It starts from 1 for $\mu=0$, is roughly $1/\mu^2$ in a crossover region 
and roughly $1/\mu$ for $\mu\gg 0$.}
\label{comparison}
\end{center}
\end{table}

\section{Conclusions}

We have found that the transport properties of a 2D electron gas with a spectral node
are controlled by a non-Abelian chiral symmetry. Spontaneous breaking of the symmetry generates massless fermion
modes that lead to diffusion: one mode if the system is at the node and two modes if the system is away
from the node. %The interaction of the two modes can be described by random diffusion coefficients. 
We have argued that the system away from the node is connected with the system at the node $\mu=0$ 
by a classical random 
walk. The latter can be approximated by a self-consistent approach. This reproduces the result that there is no
renormalization of the bare diffusion coefficient, which was recently found in a renormalization-group 
calculation of this problem \cite{sinner12}. 
%, where the interaction of the massless modes scales to zero. 

The knowledge of the diffusion coefficient enabled us to calculate the DC conductivity via the Einstein relation.
We have also seen that this is a good approximation for the Kubo formula.
For the special case of two-dimensional Dirac fermions this calculation reproduces the well-known V shape of the 
density-conductivity plot in graphene.
Therefore, our work explains the experimentally observed linear DC conductivity 
away from the charge neutrality point (Dirac node), and provides a description that is an alternative to the 
effect of charged disorder \cite{Nomura_06}.

Finally, it should be mentioned that the approximation in Eq. (\ref{expansion2}) corresponds to a weak scattering
expansion in powers of the scattering rate $\eta$. If we apply a strong scattering expansion in powers of $1/\eta$
we would see Anderson localization \cite{ziegler13}. In this regime strong fluctuations due to disorder destroy the massless
modes and create an exponentially decaying correlation function on the scale $\hbar v_F/\eta$. However, this requires 
a scattering rate of the order of the band width, which is too large to be realized for most physical systems.

In summary, we have described the transport in a two-dimensional two-band system, using spontaneous
breaking of a non-Abelian chiral symmetry. As an extension of previous calculations, our approach is applicable 
not only at the (Dirac) node but also inside the two bands. With this we found a systematic description which 
applies also to transport in graphene. In particular, we were able within a single approach to obtain (i) the 
minimal conductivity at the Dirac nodes and (ii) the V-shape conductivity inside the bands that is linear 
in the density of charge carriers.

\vskip0.2cm
\no
Acknowledgment: I am grateful to A. Sinner for interesting discussions.
This work was supported through the DFG grant ZI 305/5-1.

\appendix

%\section{Saddle-point approximation}
%\label{app:scba}

%$\eta$ satisfies the self-consistent (or saddle-point) equation for disorder strength $g$:
%\beq
%\eta=g\int_\bk\frac{z}{\lambda_1^2+\lambda_2^2+\lambda_3^2+z^2} ,\ \ \ z=\eta-i\mu
%\ , 
%\label{spe1}
%\eeq
%where $\lambda_j$ are the Fourier components of $h_j$ of the Hamiltonian. The $\bk$--integration is performed with
%respect to the Brillouin zone.

\section{coefficients of the nonlinear sigma model}
\label{app:nlsm}

The expansion terms in Eq. (\ref{expansion2}) read
\beq
{\rm Trg}\left({\hat G}_0{\hat S}^2\right)
=tr\left[(g_+ +g_-)(\varphi_1\varphi_1'+\varphi_2\varphi_2')\right]
\ ,
\label{a1}
\eeq
where $tr$ is the trace with respect to the sites and the Pauli matrices.
\beq
{\rm Trg}\left[\left({\hat G}_0{\hat S}\right)^2\right]=
{\rm Trg}\left[{\hat G}_0{\hat S}{\hat G}_0{\hat S}\right]
=2\left[ tr(g_+\varphi_1Ug_-^T\varphi_1'U^\dagger)+tr(g_-\varphi_2Ug_+^T\varphi_2'U^\dagger)
\right]
\label{a2}
\eeq
and
\beq
{\rm Trg}\left[\left({\hat G}_0{\hat S}^2\right)^2\right]=
{\rm Trg}\left[{\hat G}_0{\hat S}^2{\hat G}_0{\hat S}^2\right]
=-\sum_{j=1,2}(-1)^j
\left[
tr(g_+\varphi_j\varphi_j'g_+\varphi_j\varphi_j')-tr(g_-\varphi_j\varphi_j'g_-\varphi_j\varphi_j')
\right]
\label{a3}
\eeq
Moreover, for $\bareta=\eta+\epsilon$ we have
\beq
Ug_{\pm}(i\bareta)^T U^\dagger =-g_{\mp}(-i\bareta)
\eeq
such that we get from Eq. (\ref{a2}) 
\beq
{\rm Trg}\left[{\hat G}_0{\hat S}{\hat G}_0{\hat S}\right]
=-2tr\left[g_+(i\bareta)\varphi_1g_+(-i\bareta)\varphi_1'\right]
-2tr\left[g_-(i\bareta)\varphi_2g_-(-i\bareta)\varphi_2'\right]
\ .
\label{tr2}
\eeq
Now we assume that the Grassmann field $\varphi_{j,\br}$ varies only on large scales. This allows us to expand the expressions
(\ref{a1})--(\ref{a3}) in powers of the wave vector $\bq$ up to second order.
Then we have, together with the property $U U^\dagger={\bf 1}$,
\[
tr\left[g_\pm(i\bareta)\right]=-tr\left[g_\mp(-i\bareta)\right] ,  \ \ \
tr\left[g_+(i\bareta)+g_-(i\bareta)\right]=tr\left[g_+(i\bareta)-g_+(-i\bareta)\right]
\ .
\]
and
\[
tr\left[g_\pm(i\bareta)-g_\pm(-i\bareta)\right]=-2i\bareta tr\left[g_\pm(i\bareta)g_\pm(-i\bareta)\right]
\]
such that
\[
tr\left[g_+(i\bareta)+g_-(i\bareta)\right]=-2i\bareta tr\left[g_\pm(i\bareta)g_\pm(-i\bareta)\right]
\ .
\]
This gives us Eq. (\ref{action4}) with the coefficients (\ref{coeff1}), (\ref{coeff2}).

\section{Self-consistent calculation of the scattering rate}
\label{app:b}

%\beq
%\langle G(z)\rangle\approx G_0(z+i\eta) , \ \ \ G_0(z)=(\langle H\rangle -z)^{-1}
%\ ,
%\label{scba}
%\eeq
%where the self-energy $\eta$ is a scattering rate, which is determined by the self-consistent equation
%$i\eta= G_{0,0}(z+i\eta)$ \cite{mahan}. 

The scattering rate $\eta=1/\tau$ is a function of the Fermi energy $\mu$ and can be evaluated within the 
self-consistent equation $\eta= 2ig G_{0,0}(z+i\eta)$ which reads for small $g$
\beq
\frac{g\nu}{\mu}=\frac{e^{-\kappa\tan \kappa}\cos\kappa}{\kappa} \ \ \ (-\pi/2<\kappa<\pi/2)
\ \ \ \nu=\lambda e^{-1/g}
\ .
\label{spe3}
\eeq
This equation determines $\kappa$ and the latter gives
\beq
\eta=\nu e^{\kappa\tan\kappa}\cos\kappa , \ \ \  
\frac{\bar\mu}{\eta}=\frac{g\kappa+\cos\kappa\sin\kappa}{\cos^2\kappa} %\ \ \ (\nu=\lambda e^{-1/g})
\ .
\label{parameters}
\eeq

\section{Random walk expansion}
\label{app:c}

The random walk expansion is a convenient tool to calculate the integral of Eq. (\ref{corra}) \cite{glimm}.
In our case it leads to a perturbation series in powers of $\alpha_j$. It is be based on the 
expansion of $\exp(-S_j'')$ with $S_j''$ given in Eq. (\ref{action4}):
\beq
\exp(-S_j'')=\exp\left(\sum_\br\varphi_{j\br}(\epsilon -D\partial^2)\varphi_{j\br}'\right)
\prod_\br\left(1+\alpha_j\varphi_{j\br}\varphi_{j\br}'\partial^2\varphi_{j\br}\varphi_{j\br}'\right)
.
\label{exp01}
\eeq
For this result and for the following discussion it is crucial to keep in mind that Grasmann variables
are nilpotent (i.e., $\varphi_{j\br}^n=\varphi_{j\br}'^n=0$ for integer $n>1$). 
From the definition $\partial_j\Phi_\br=(\Phi_{\br+a e_j}-\Phi_{\br-a e_j})/a$ we get
$\partial_j^2\Phi_\br=(\Phi_{\br+2a e_j}+\Phi_{\br-2a e_j}-2\Phi_{\br})/a^2$.
First we notice that
\[
\varphi_{j\br}\varphi_{j\br}'\partial^2\varphi_{j\br}\varphi_{j\br}'
=-\varphi_{j\br}\varphi_{j\br}'\frac{1}{a^2}\sum_{n=1,2}(\varphi_{j\br+2a e_n}
\varphi_{j\br+2a e_n}'+\varphi_{j\br-2a e_n}\varphi_{j\br-2a e_n}')
\ .
\]
%where we have used that the Grassmann variables are nilpotent. 
Thus, the second factor in (\ref{exp01})
provides at each pair of sites (dimer) $\br,\br\pm 2a e_n$ either a factor 1 or a product of Grassmann variables 
$\varphi_{j\br}\varphi_{j\br}'\varphi_{j\br\pm 2a e_n}\varphi_{j\br\pm 2a e_n}'$,
%$\varphi_{j\br}\varphi_{j\br}'\varphi_{j\br-2a e_n}\varphi_{j\br-2a e_n}'$,
respectively. This allows us to write
\beq
\exp(-S_j'')=\exp\left(\sum_{\br\in\Lambda}\varphi_{j\br}(\epsilon -D\partial^2)\varphi_{j\br}'\right)
\sum_{I}{\prod}'_{\br,\br'\in I} \frac{\alpha_j}{a^2}\varphi_{j\br}\varphi_{j\br}'\varphi_{j\br'}\varphi_{j\br'}'
\ ,
\label{exp02}
\eeq
where the product $\prod_{\br,\br'\in I}'$ is restricted to dimers $\br'=\br\pm 2ae_n$ 
and $I\subseteq \Lambda$ is a subset of our real space $\Lambda$. 
Next we can expand the first factor in (\ref{exp01}) by using
\[
\sum_{\br\in\Lambda}\varphi_{j\br}(\epsilon -D\partial^2)\varphi_{j\br}'
=(\epsilon+2D_a)\sum_{\br\in\Lambda}\varphi_{j\br}\varphi_{j\br}'
-D_a\sum_{\br\in\Lambda}\sum_{n=1,2}\varphi_{j\br}(
\varphi_{j\br+2a e_n}'+\varphi_{j\br-2a e_n}') \ \ \ (D_a=D/a^2)
\]
such that
\[
\exp\left(\sum_{\br\in\Lambda}\varphi_{j\br}(\epsilon -D\partial^2)\varphi_{j\br}'\right)
%=\exp\left((\epsilon+2D_a)\sum_{\br\in\Lambda}\varphi_{j\br}\varphi_{j\br}'\right)\prod_{\br,\br'}
%\left(1-D_a\varphi_{j\br}\varphi_{j\br'}'\right) 
%\]
%\[
=\exp\left((\epsilon+2D_a)\sum_{\br\in\Lambda}\varphi_{j\br}\varphi_{j\br}'\right)\sum_{I'}{\prod}'_{\br,\br'\in I'}
\left(-D_a\varphi_{j\br}\varphi_{j\br'}'\right) 
\]
Combining the expansion of the two factors in Eq. (\ref{exp01}) gives us
\beq
\exp(-S_j'')=
%\exp\left((\epsilon+2D_a)\sum_{\br\in\Lambda}\varphi_{j\br}\varphi_{j\br}'\right)
\sum_{I,I'}\prod_{\br\in\Lambda}[1+(\epsilon+2D_a)\varphi_{j\br}\varphi_{j\br}']
{\prod}'_{\br,\br'\in I} \left(-D_a\varphi_{j\br}\varphi_{j\br'}'\right) 
{\prod}'_{\br,\br'\in I'} \frac{\alpha_j}{a^2}\varphi_{j\br}\varphi_{j\br}'\varphi_{j\br'}\varphi_{j\br'}'
\ .
\label{exp03}
\eeq
$I$ and $I'$ must be disjunct (i.e., they have no common sites $\br$) to give a non-zero contribution to the sum
due to the nilpotent Grassmann variables. Thus, the approximated Jacobian $\exp(-S_j'')$ consists of (I) factors 
$\frac{\alpha_j}{a^2}\varphi_{j\br}\varphi_{j\br}'\varphi_{j\br'}\varphi_{j\br'}'$ from dimers $\br,\br'\in I'$,
(II) factors $-D_a\varphi_{j\br}\varphi_{j\br'}'$ from dimers $\br,\br'\in I$ and (III) factors 
$(\epsilon+2D_a)\varphi_{j\br}\varphi_{j\br}'$ on points $\br$, which do neither belong to $I'$ nor to $I$. 
Inserting this result into the correlation function of Eq. (\ref{corra}), we can perform the integration over the Grassmann variables for each
term of the sums in Eq. (\ref{exp03}). At each site $\br$ the integral gives 1 if there is a factor $\varphi_{j\br}\varphi_{j\br}'$ and
zero otherwise. Then the result of the expansion can be represented in a graphical manner: Since there are only isolated points (with $(\epsilon+2D_a)$),
isolated dimers (with $\alpha_j/a^2$) or connected dimers (with $-D_a$) after the Grassmann integration, we can depict these three types
of elements as isolated points, isolated dimers and connected dimers in space, respectively. We also have to take into account the extra Grassmann variables
in the integrand $\varphi_{j\br} \varphi'_{j\br'}$. They can only appear as end points of a chain of connected dimers, as it is shown in Fig. \ref{fig:dimers}.
The other connected dimers form closed loops and cannot intersect with each other. 

Eq. (\ref{exp03}) is identical to
\[
\exp(-S_j'')=
\sum_{I}\prod_{\br\in I}[1+(\epsilon+2D_a)\varphi_{j\br}\varphi_{j\br}']
{\prod}'_{\br,\br'\in I} \left(1-D_a\varphi_{j\br}\varphi_{j\br'}'\right) 
{\prod}'_{\br,\br'\in \Lambda\backslash I} \frac{\alpha_j}{a^2}\varphi_{j\br}\varphi_{j\br}'\varphi_{j\br'}\varphi_{j\br'}'
\]
\beq
=\sum_{I}\exp\left(\sum_{\br\in I}\varphi_{j\br}(\epsilon -D\partial^2)\varphi_{j\br}'\right)
{\prod}'_{\br,\br'\in \Lambda\backslash I} \frac{\alpha_j}{a^2}\varphi_{j\br}\varphi_{j\br}'\varphi_{j\br'}\varphi_{j\br'}'
\ .
\label{exp04}
\eeq
Inserting this into Eq. (\ref{corra}) and performing the integration with respect to the Grassmann variables
gives us Eq. (\ref{pert_exp}).


\begin{thebibliography}{99}

\bibitem{novoselov05}
K.S. Novoselov, A.K. Geim, S.V. Morozov, D. Jiang, M.I. Katsnelson, I.V. Grigorieva, S.V. Dubonos,
A.A. Firsov, Nature {\bf 438}, 197 (2005).

\bibitem{zhang05}
Y. Zhang, Y.-W. Tan, H.L. Stormer, P. Kim, Nature {\bf 438}, 201 (2005).

\bibitem{castro09}
A.H. Castro Neto, F. Guinea, N.M.R. Peres, K.S. Novoselov, and A.K. Geim,
Rev. Mod. Phys. {\bf 81}, 109 (2009).

\bibitem{abergel10}
D.S.L. Abergel, V. Apalkov, J. Berashevich, K. Ziegler and T. Chakraborty, Adv. Phys. {\bf 59}, 261  (2010).

\bibitem{zhang11}
X.-L. Qi and S.-C. Zhang, Rev. Mod. Phys. {\bf 83}, 1057 (2011).

\bibitem{abrahams79}
E. Abrahams, P.W. Anderson, D.C. Licciardello and T.V. Ramakrishnan, Phys. Rev. Lett. {\bf 42}, 673 (1979).

\bibitem{altshuler}
B.L. Altshuler et al., Phys. Rev. B {\bf 22}, 5142 (1980);
B.L. Altshuler and B.D. Simons, in {\it Mesoscopic quantum physics},
eds. E. Akkermans et al. (North-Holland 1995).

\bibitem{andoetal}
T. Ando, Y. Zheng and H. Suzuura, J. Phys. Soc. Japan {\bf 71}, 1318 (2002);
H. Suzuura and T. Ando, Phys. Rev. Lett. {\bf 89}, 266603 (2002); E. McCann et al.,
Phys. Rev. Lett. {\bf 97}, 146805 (2006).

\bibitem{khveshchenko06}
D.V. Khveshchenko, Phys. Rev. Lett., {\bf 97}, 036802 (2006).

\bibitem{langer66}
J.S. Langer and T. Neal, Phys. Rev. Lett. {\bf 16}, 984 (1966).

\bibitem{rosenstein09}
M. Lewkowicz and B. Rosenstein, Phys. Rev. Lett. {\bf 102}, 106802 (2009).

\bibitem{economou70}
E.N. Economou and M.H. Cohen, Phys. Rev. Lett. {\bf 25} (1970).

\bibitem{wegner80}
L. Sch\"afer and F. Wegner, Z. Physik B {\bf 38}, 113 (1980).

\bibitem{efetov97}
K. Efetov, {\it Supersymmetry in Disorder and Chaos} (Cambridge University Press 1997).

\bibitem{bocquet00}
M. Bocquet, D. Serban and M.R. Zirnbauer, Nucl. Phys. B {\bf 578}, 628 (2000).

\bibitem{ziegler97}
K. Ziegler, Phys. Rev. B {\bf 55}, 10661 (1997); Phys. Rev. Lett. {\bf 80}, 3113 (1998).

\bibitem{zirnbauer96}
M.R. Zirnbauer, J. Math. Phys. {\bf 37}, 4986 (1996); A. Altland and M.R. Zirnbauer, Phys. Rev. B {\bf 55}, 1142 (1997).
%; Nucl. Phys. B 578, 628 (2000); Phys. Rep. 359, 283 (2002); 
%Rev. Mod. Phys. 80, 1355 (2008).

\bibitem{Ziegler2009} 
K. Ziegler, Phys. Rev. Lett. {\bf 102}, 126802 (2009); Phys. Rev. B {\bf 79}, 195424 (2009).

\bibitem{senthil00}
T. Senthil and M.P.A. Fisher, Phys. Rev. B {\bf 61}, 9690 (2000). 

\bibitem{chalker01}
J.T. Chalker, N. Read, V. Kagalovsky, B. Horovitz, Y. Avishai, A.W.W. Ludwig,
Phys. Rev. B {\bf 65} 012506 (2001).

\bibitem{evers08}
F. Evers and A.D. Mirlin, Rev. Mod. Phys. {\bf 80}, 1355 (2008).

\bibitem{beenakker10}
M.V. Medvedyeva, J. Tworzydlo, and C.W.J. Beenakker,
Phys. Rev. B {\bf 81}, 214203 (2010).

\bibitem{coleman}
S. Coleman, {\it Aspects of symmetry} (Cambridge University Press 1985).

\bibitem{ando98}
N.H. Shon and T. Ando, J. Phys. Soc. Japan {\bf 67}, 2421 (1998).

\bibitem{ziegler12}
K. Ziegler, J. Phys. A: Math. Theor. {\bf 45}, 335001 (2012).

\bibitem{sinner12}
A. Sinner and K. Ziegler, Phys. Rev. B {\bf 86}, 155450 (2012).

\bibitem{bernad10}
J.Z. Bern\'ad, U. Z\"ulicke and K. Ziegler, Physica E {\bf 42}, 755 (2010).

\bibitem{chen08}
J. H. Chen, C. Jang, M. S. Fuhrer, E. D. Williams, M. Ishigami,
Nature Physics {\bf 4}, 377 (2008).

\bibitem{andrei12}
E.Y. Andrei, G. Li  and X. Du, 
Rep. Prog. Phys.,  {\bf 75}, 056501 (2012).


%\bibitem{ando02}
%H. Suzuura and T. Ando, Phys. Rev. Lett. {\bf 89}, 266603 (2002).

\bibitem{sinner11}
A. Sinner and K. Ziegler, Phys. Rev. B {\bf 84}, 233401 (2011).

\bibitem{Nomura_06}
K. Nomura and A.H. MacDonald, Phys. Rev. Lett. {\bf 96}, 256602 (2006).

\bibitem{ziegler13}
A. Hill and K. Ziegler, arXiv:1305.6901.

\bibitem{glimm}
J. Glimm and A. Jaffe, {\it Quantum Physics} (Springer-Verlag 1981).


\end{thebibliography}
\end{document}